\documentclass[aps,prc,onecolumn,11pt,groupedaddress]{revtex4-1}
\usepackage{amsmath}
\usepackage{bm}
\usepackage{graphicx}
\usepackage{wasysym}
\usepackage{amssymb}

\begin{document}
\title{Triaxiality induced monopole-quadrupole-hexadecupole coupling in the isoscalar giant resonances for $^{86}\textrm{Ge}$}
\author{Xuwei Sun}
\affiliation{State Key Laboratory of Nuclear Physics and Technology, School of Physics, Peking University, Beijing 100871, China}
\author{Jie Meng}
\email{mengj@pku.edu.cn}
\affiliation{State Key Laboratory of Nuclear Physics and Technology, School of Physics, Peking University, Beijing 100871, China}
\affiliation{Yukawa Institute for Theoretical Physics, Kyoto University, Kyoto 606-8502, Japan}

\begin{abstract}
The isoscalar giant resonances for $^{86}\textrm{Ge}$ are studied by the quasiparticle finite amplitude method based on the covariant density functional theory. 
In addition to the well-known monopole-quadrupole coupling that splits the isoscalar giant monopole resonance in axially deformed nuclei, 
a monopole-quadrupole-hexadecupole coupling is identified in the triaxially deformed nucleus $^{86}\textrm{Ge}$,
leading to the emergence of a distinct resonance peak at the low energy side of the isoscalar monopole strength function.
The transition density of the triaxiality induced resonance peak shows a strong interplay among monopole, quadrupole, and hexadecupole vibrations.
The resonance peak responses to monopole, quadrupole, and hexadecupole perturbations simultaneously, which could be regarded as a fingerprint of the triaxiality in $^{86}\textrm{Ge}$.
\end{abstract}

\maketitle

\section{Introduction}

Giant resonances (GRs) \cite{Book_Harajeh} are small amplitude collective vibrations of a nucleus, which are related to nuclear bulk properties, e.g., via certain sum rules \cite{Lipparini_PR_1989}, and provide valuable information on nuclear structure such as incompressibility \cite{Blaizot_PR_1980,Li_PRL_2007,Giai_NPA_1981,Youngblood_PRL_1999,Garg_PPNP_2018}, symmetry energy  \cite{Trippa_PRC_2008,Cao_PRC_2015,Roca_PRC_2013}, and neutron skin thickness \cite{Klimkiewicz_PRC_2007,Carbone_PRC_2010,Inakura_PRC_2011,Sun_CPC_2018}.

The effect of axial deformation describing prolate or oblate spheroid shapes on GRs has been extensively discussed \cite{Gupta_PLB_2015,Itoh_PLB_2002,Peach_PRC_2016}.
In particular, shortly after the discovery of the isoscalar giant monopole resonance (ISGMR) \cite{Harakeh_PRL_1977}, it had been reported that the ISGMRs for well-deformed nuclei split into two branches \cite{Garg_PRL_1980}. 
The splitting is caused by the coupling between ISGMR and isoscalar giant quadrupole resonance (ISGQR) with $K=0$, namely, the monopole-quadrupole coupling (E0-E2 coupling). Experimental evidences of the E0-E2 coupling are found for $^{150}\textrm{Nd}$ \cite{Garg_PRC_1984}, $^{154}\textrm{Sm}$ \cite{Youngblood_PRC_1999}, $^{181}\textrm{Ta}$ \cite{Buenerd_PRL_1980}, $^{238}\textrm{U}$ \cite{Brandenburg_PRL_1982}, etc.
Explanations and predictions of the E0-E2 coupling are made by macroscopic calculations through cranking model \cite{Abgrall_NPA_1980}, variational procedure \cite{Jang_NPA_1983}, fluid dynamical description \cite{Nishizaki_PTP_1985}, and random phase approximation calculations \cite{Niksic_PRC_2013,Kvasil_PRC_2016,Colo_PLB_2020}.

Apart from elongating or compressing along the symmetry axis of the intrinsic frame, a nucleus can be further squeezed perpendicularly to its symmetry axis, and has a triaxially deformed shape. 
Triaxial deformation is thought to be the key ingredient for a lot of interesting phenomena such as nuclear chirality \cite{Frauendorf_NPA_1997,Starosta_PRL_2001} and wobbling \cite{Odegard_PRL_2001}.
Triaxiality will also affect GRs in a significant way.
For example, a recent work suggests that triaxiality sheds light upon the softening of ISGMRs for cadmium isotopes \cite{Sun_PRC_2019_2}.

The microscopic random phase approximation (RPA) method \cite{book_RingSchuck} is widely used in studying nuclear collective vibrations. Combined with modern nuclear energy density functional \cite{Niksic_PPNP_2011,book_Meng,Stone_PPNP_2007}, RPA method is able to give successful descriptions to nuclear giant resonances \cite{Sun_PRC_2019,Gupta_PRC_2016,Sun_PRC_2018}. 
For deformed and super-fluid nuclei, the full configuration space of RPA is huge and solving RPA equation is extremely challenging
if no artificial truncation is applied. 
Finite amplitude method (FAM) \cite{Nakatsukasa_PRC_2007} provides a numerical feasible way to solve large scale RPA problems. In spherical and axially deformed cases, FAM has been implemented both on relativistic \cite{Liang_PRC_2013,Sun_PRC_2017,Niksic_PRC_2013,Bjelcic_CPC_2020,Sun_PRC_2021} and non-relativistic \cite{Inakura_PRC_2009,Avogadro_PRC_2011,Mustonen_PRC_2014,Kortelainen_PRC_2015} density functionals. In triaxially deformed case, FAM is so far implemented for Skyrme density functional theory in the three-dimensional Cartesian coordinate space \cite{Washiyama_PRC_2017}, as well as for covariant density functional theory (CDFT) in a triaxially deformed harmonic oscillator basis \cite{Sun_PRC_2019_2}, making it possible to study triaxial deformation effects on GRs in a microscopic and self-consistent way.

Though the E0-E2 coupling in axially deformed nuclei has been well understood, the situation in triaxially deformed nuclei demands further study. It is also interesting to search for possible distinctive resonance structures caused by the triaxiality. 
The paper is devoted to investigate the impact of triaxiality on the GRs by the triaxially deformed quasiparticle finite amplitude method (QFAM) on CDFT \cite{Sun_PRC_2019_2}.
The outline of the paper is as follows. Section II presents the formalism of CDFT and QFAM. In Section III, the triaxiality effects on the isoscalar giant resonances for the triaxially deformed nucleus $^{86}\textrm{Ge}$ will be analyzed. Conclusions and remarks will be given in Section IV.

\section{Formalism}
In this section, the nuclear CDFT \cite{book_Meng} with effective meson-exchange interaction, and the implementation of the QFAM will be briefly introduced.

In CDFT, the effective nuclear forces are mediated by the scalar meson $\sigma$, the vector meson $\omega_{\mu}$, the vector-isovector meson $\vec{\rho}_{\mu}$, and the photon $A_{\mu}$. The interacting part of the Lagrangian density is
\begin{equation}
	\mathcal{L}_{\textrm{int}}=-\sum_{m}g_{m}\bar{\psi}\Gamma_m\cdot\phi_{m}\psi,
\end{equation}
where $\Gamma_m=\{1,\gamma^{\mu},\vec{\tau}\gamma^{\mu},(1-\tau_3)\gamma^{\mu}/2\}$ is the coupling vertex between the nucleon $\psi$ and the meson (photon) $\phi_{m}=\{\sigma,\omega_{\mu},\vec{\rho}_{\mu},A_{\mu}\}$ with the coupling strength $g_m$.

The state of a nuclear system $|\Phi\rangle$ can be uniquely expressed by the density operator $\hat{\rho}$. The matrix element of density operator $\hat{\rho}$ in a single particle basis has the form \cite{book_RingSchuck}
\begin{equation}
\rho_{pq}=\langle\Phi|c_q^\dagger c_p|\Phi\rangle.
\end{equation}
The Hamiltonian density can be transformed from the Lagrangian density, its expectation value in the state $|\Phi\rangle$ is the following density functional
\begin{equation}
\begin{aligned}
\epsilon[\hat{\rho},\phi]
=&\textrm{Tr}[(-i\bm{\alpha}\nabla+\beta m)\hat{\rho}]
+\sum_m\textrm{Tr}[\beta g_m\Gamma_m\phi_m\hat{\rho}]\\
&\pm\sum_m\frac12\int\!d^3r[(\partial_{\mu}\phi_m)^2+m_m^2\phi_m^2].\\
\end{aligned}
\end{equation}
The equation of motion of the density operator and meson (photon) field can be derived from the time-dependent variation principle
\begin{equation}
\delta\int_{t_1}^{t_2}dt\{\langle\Phi|i\partial_t|\Phi\rangle-\epsilon[\hat{\rho},\phi]\}=0,
\end{equation}
which reads
\begin{equation}
\begin{aligned}
i\partial_t\hat{\rho}&=[\hat{h},\hat{\rho}],\\
(\partial^{\nu}\partial_{\nu}+m_m^2)\phi_m&=\mp\textrm{Tr}[\beta g_m\Gamma_m\hat{\rho}].\\
\end{aligned}
\end{equation}
The single particle Hamiltonian $\hat{h}$ is
\begin{equation}\label{hd}
\hat{h}[\hat{\rho},\phi]\equiv\frac{\delta\epsilon[\hat{\rho},\phi]}{\delta\hat\rho}
=-\bm{\alpha}(i\nabla+\bm{V})+V_0+\beta(m+S),
\end{equation}
with the scalar potential $S$ and the vector potential $V_{\mu}\equiv(V_0,\bm{V})$ consisting of the meson and photon fields,
\begin{equation}
\begin{aligned}
S&=g_{\sigma}\sigma,\\
V_{\mu}&=g_{\omega}\omega_{\mu}+g_{\rho}\vec{\tau}\cdot\vec{\rho}_{\mu}+e\frac{1-\tau_3}{2}A_{\mu}
+\Sigma_{\mu}^{R}.\\
\end{aligned}
\end{equation}
The rearrangement term $\Sigma_{\mu}^{R}$ appears when the coupling strength $g_m$ is density-dependent \cite{Lalazissis_PRC_2005}.

With a pairing interaction $V^{pp}$, the matrix elements of the pairing tensor $\hat{\kappa}$ is
\begin{equation}
\kappa_{pq}=\langle\Phi|c_q c_p|\Phi\rangle,
\end{equation}
from which the pairing energy is evaluated, 
\begin{equation}
\varepsilon[\hat{\kappa}]=\frac{1}{4}\mathrm{Tr}[\hat{\kappa}^*V^{pp}\hat{\kappa}],
\end{equation}
and a pairing potential $\hat{\Delta}$ can be defined:
\begin{equation}
	\hat{\Delta}\equiv\frac{\delta\varepsilon[\hat{\kappa}]}{\delta\hat{\kappa}}.
\end{equation}
The density and the paring tensor can be expressed in a compact form via the generalized density \cite{Valatin_PR_1961}
\begin{equation}
\hat{\mathcal{R}}=
\left(\!
\begin{array}{cc}
\hat{\rho}&\hat{\kappa}\\
-\hat{\kappa}^*&1-\hat{\rho}^*\\
\end{array}
\!\right),
\end{equation}
and the generalized Hamiltonian is obtained accordingly
\begin{equation}
\hat{\mathcal{H}}= \frac{
%\delta E[\hat{\mathcal{R}}]
\delta(\epsilon[\hat{\rho},\phi]+\varepsilon[\hat{\kappa}])
}{\delta\hat{\mathcal{R}}}=
\left(\!
\begin{array}{cc}
\hat{h}&\hat{\Delta}\\
-\hat{\Delta}^*&-\hat{h}^*\\
\end{array}
\!\right).
\end{equation}
Diagonalizing the generalized Hamiltonian gives the relativistic Hartree-Bogoliubov (RHB) equation \cite{book_RingSchuck}
\begin{equation}\label{eqrhb}
	\bigg(\!
	\begin{array}{cc}
		\hat{h}-\lambda&\hat{\Delta}\\
		-\hat{\Delta}^*&-\hat{h}^*+\lambda\\
	\end{array}
	\!\!\bigg)
	\bigg(\!
	\begin{array}{c}
		U_{\mu}\\
		V_{\mu}\\
	\end{array}
	\!\!\bigg)
	=E_{\mu}\bigg(\!\!
	\begin{array}{c}
		U_{\mu}\\
		V_{\mu}\\
	\end{array}
	\!\!\bigg),
\end{equation}
where a chemical potential $\lambda$ is introduced to account for particle number conservation \cite{book_RingSchuck}. $E_{\mu}$ is the quasiparticle energy. The RHB equation is solved by an expansion of the quasiparticle spinor $U_{\mu}$, $V_{\mu}$ on a triaxial harmonic oscillator basis \cite{Niksic_CPC_2014}, which is the product of three one-dimensional harmonic oscillator wavefunction and a spin factor,
\begin{equation}
\phi_i(x,y,z) = \varphi_{n_x}(x)\varphi_{n_y}(y)\varphi_{n_z}(z)\chi_{m_s},
\end{equation}
labeled by a set of quantum number $i \equiv\{n_x,n_y,n_z;m_s\}$.

If a nucleus is perturbed by a weak external field $\mathcal{F}(t)$ with the frequency $\omega$, its density $\mathcal{R}(t)$ and Hamiltonian $\mathcal{H}(t)$ vibrates slightly around the equilibrium $\mathcal{R}_0$ and $\mathcal{H}_0$. The following small-amplitude approximation is valid,
\begin{equation}\label{small_amp}
	\begin{aligned}
		&\mathcal{R}(t)=\mathcal{R}_0+\delta R(\omega)e^{-i\omega t}+\text{H.c.},\\
		&\mathcal{H}(t)=\mathcal{H}_0+\delta H(\omega)e^{-i\omega t}+\text{H.c.}.\\
	\end{aligned}
\end{equation}
From the equation of motion 
$i\dot{\mathcal{R}}(t)=[\mathcal{H}(t)+\mathcal{F}(t),\mathcal{R}(t)]$,
the linear response equation can be obtained,
\begin{equation}\label{fameq}
	\begin{aligned}
		&(E_{\mu}+E_{\nu}-\omega)X_{\mu\nu}(\omega)+\delta H_{\mu\nu}^{20}(\omega)= -F_{\mu\nu}^{20},\\
		&(E_{\mu}+E_{\nu}+\omega)Y_{\mu\nu}(\omega)+\delta H_{\mu\nu}^{02}(\omega)= -F_{\mu\nu}^{02}.\\
	\end{aligned}
\end{equation}
The quasiparticle energy $E_{\mu}$, $E_{\nu}$, and the matrix element of external field $F_{\mu\nu}^{20}$, $F_{\mu\nu}^{02}$ are independent to the excitation energy $\omega$.
The induced Hamiltonian $\delta H(\omega)$ is a functional of the transition amplitudes $X(\omega)$ and $Y(\omega)$, they are calculated via the following finite amplitude method.

The matrix elements of the induced Hamiltonian $\delta H(\omega)$ in quasiparticle basis can be calculated from the variation of the single-particle Hamiltonian $\delta h(\omega)$ and the variation of the pairing potential $\delta\Delta(\omega)$, $\delta\Delta^*(\omega)$,
\begin{equation}\label{vh}
	\begin{aligned}
		\delta H^{20}_{\mu\nu}(\omega)=& [U^{\dagger}\delta h(\omega) V^* - V^{\dagger}\delta h^T(\omega) U^*\\
		                                             &-V^{\dagger}\delta\Delta^*(\omega)V^*+ U^{\dagger}\delta\Delta(\omega) U^*]_{\mu\nu},\\
		\delta H^{02}_{\mu\nu}(\omega)=&[U^T \delta h^T(\omega) V -V^T \delta h(\omega) U\\
		                                             &-V^T \delta\Delta(\omega) V +U^T \delta\Delta^*(\omega) U]_{\mu\nu}.\\
	\end{aligned}
\end{equation}
The transition amplitude $X(\omega)$ and $Y(\omega)$ are used to get the variation of the single-particle density $\delta\rho(\omega)$ and the variation of the pairing tensor $\delta\kappa(\omega)$, $\delta\kappa^*(\omega)$, whose matrix elements in the harmonic oscillator basis read
\begin{equation}\label{vd}
	\begin{aligned}
		\delta\rho_{ij}(\omega) &= [UX(\omega)V^T+V^*Y(\omega)U^{\dagger}]_{ij},\\
		\delta\kappa_{ij}(\omega) &= [UX(\omega)U^T+V^*Y(\omega)V^{\dagger}]_{ij},\\
		\delta\kappa^*_{ij}(\omega) &=-[VX(\omega)V^T-U^*Y(\omega)V^{\dagger}]_{ij}.\\
	\end{aligned}
\end{equation}

The variation $\delta h(\omega)$ ($\delta\Delta(\omega)$ and $\delta\Delta^*(\omega)$) are calculated from the difference between the single-particle Hamiltonian (the pairing field) at the perturbed density and at the equilibrium $\rho_0$  ($\kappa_0$ and $\kappa_0^*$),
\begin{equation}\label{num_diff}
	\begin{aligned}
		\delta h(\omega)&=\frac{1}{\eta}(h[\rho_0+\eta\delta\rho(\omega)]-h[\rho_0]),\\
		\delta \Delta(\omega)&=\frac{1}{\eta}(\Delta[\kappa_0+\eta\delta\kappa(\omega)]-\Delta[\kappa_0]),\\
		\delta \Delta^*(\omega)&=\frac{1}{\eta}(\Delta^*[\kappa_0^*+\eta\delta\kappa^*(\omega)]-\Delta^*[\kappa_0^*]),\\
	\end{aligned}
\end{equation}
with a small real parameter $\eta=10^{-6}$.

Once the connection from the transition amplitude to the induced Hamiltonian is established,
the Eq. \eqref{fameq} can be solved iteratively.
The converged transition amplitudes $X_{\mu\nu}(\omega)$ and $Y_{\mu\nu}(\omega)$ are used to calculate the strength function,
\begin{equation}
	S_F(\hat{F},\omega)
	=-\frac{1}{\pi}\mathrm{Im}\sum_{\mu\nu}\{F^{20*}_{\mu\nu}X_{\mu\nu}(\omega)
	+F^{02*}_{\mu\nu}Y_{\mu\nu}(\omega)\}.\\
\end{equation}

\begin{figure}	
	\centering
	%\includegraphics[width=0.3\textwidth]{./Figure_1.eps}\\		
	\includegraphics[width=0.5\textwidth]{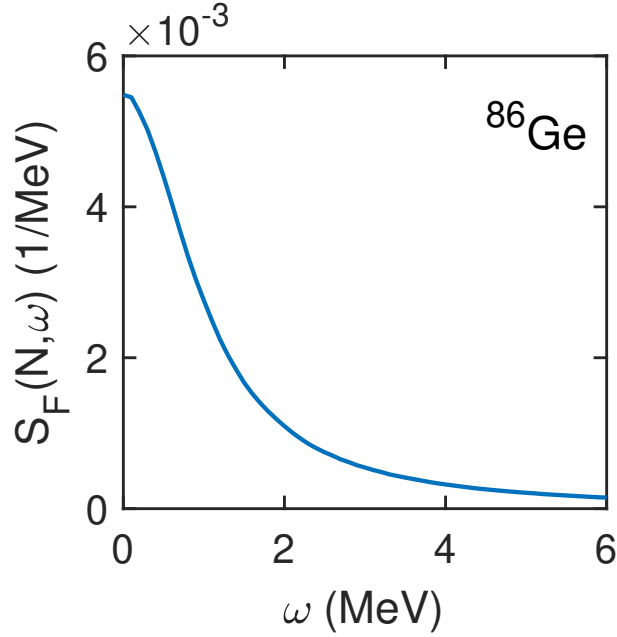}\\		
	\caption{Strength function for the particle number operator $\hat{N}$. The spurious state centers at zero energy, since the particle-number conservation has been restored by the QFAM calculation.\label{Ge86_SP}}
\end{figure}
In this work, the used functional is DD-ME2 \cite{Lalazissis_PRC_2005} and the pairing interaction is a separable pairing force \cite{Tian_PLB_2009}. 
Due to the numerical feasibility of QFAM, there is no artificial truncation being applied on the two quasiparticle pairs. 
The calculations are performed on a harmonic oscillator basis with 12 shells, the amount of the two quasiparticle pairs involved in the QFAM calculation is $N_{2qp} = 2,086,560$.
In fully self-consistent QFAM calculations, the symmetry broken in the ground state will be restored, which automatically generates a zero-energy Nambu-Goldstone boson. For instance, the particle number conservation broken in the RHB level is recovered in the QFAM calculation. Therefore, a spurious state with a vanishing energy appears in the strength function for the particle number operator $\hat{N}$, as illustrated in Fig. \ref{Ge86_SP}.

\section{Results and discussions}
\begin{figure}	
	\centering
	%\includegraphics[width=0.35\textwidth]{./Figure_2.eps}\\		
	\includegraphics[width=0.7\textwidth]{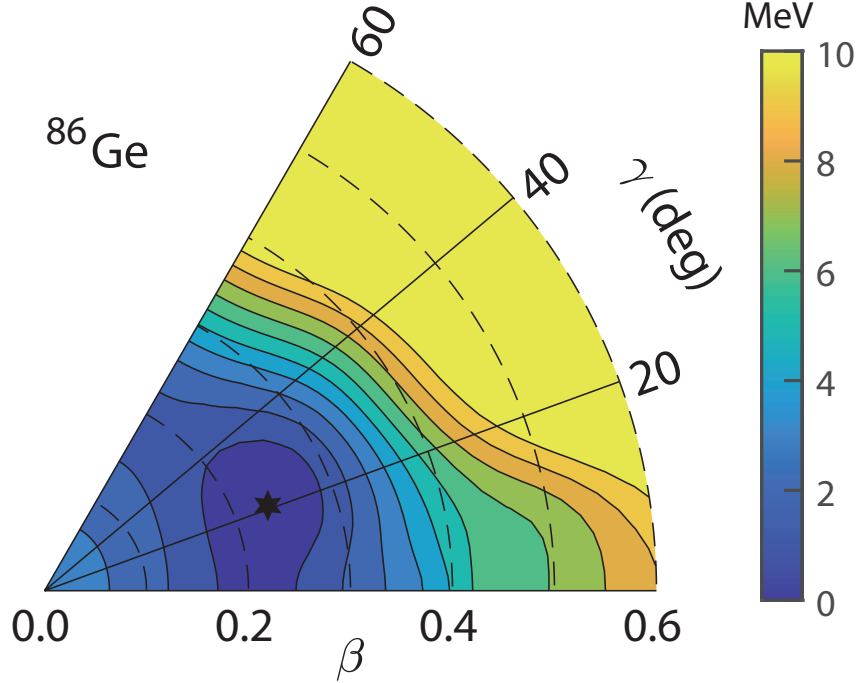}\\		
	\caption{Potential-energy surface of $^{86}\textrm{Ge}$ calculated by CDFT with DD-ME2 and a separable pairing force. The contours are drawn with a spacing of 1 MeV. The minimal of the PES is presented with a hexagram.\label{Ge86_PES}}
\end{figure}
The potential-energy surface (PES) with the deformation parameter $(\beta,\gamma)$ \cite{Niksic_CPC_2014} is illustrated in Fig. \ref{Ge86_PES} for $^{86}\textrm{Ge}$. The PES shows a minimal at $\beta = 0.234$ and $\gamma = 20.9^{\circ}$, namely, the ground state of $^{86}\textrm{Ge}$ is triaxially deformed. 
In axially deformed calculation, there are a prolate minimal at $\beta = 0.209$ with $E=0.62$ and an oblate minimal at $\beta=0.191$ with $E=1.70$ MeV, the ground state is prolate deformed.
Compared to the axially deformed case, the existence of triaxial deformation makes $^{86}\textrm{Ge}$ more elongated in $z$ direction.

The lifetime of the nucleus $^{86}\textrm{Ge}$ is about 226 ms \cite{Negret_NDS_2015}, which might make giant resonances possible with the rapid development of radioactive beam facilities. 
In order to look for possible vibration properties peculiar to triaxiality, 
both triaxial QFAM calculation and axial QFAM calculation are preformed to study the giant resonances for $^{86}\textrm{Ge}$, with the density functional DD-ME2 \cite{Lalazissis_PRC_2005} and the separable pairing force  \cite{Tian_PLB_2009}.
\begin{figure}	
	\centering
	%\includegraphics[width=0.45\textwidth]{./Figure_3.eps}\\
	\includegraphics[width=0.7\textwidth]{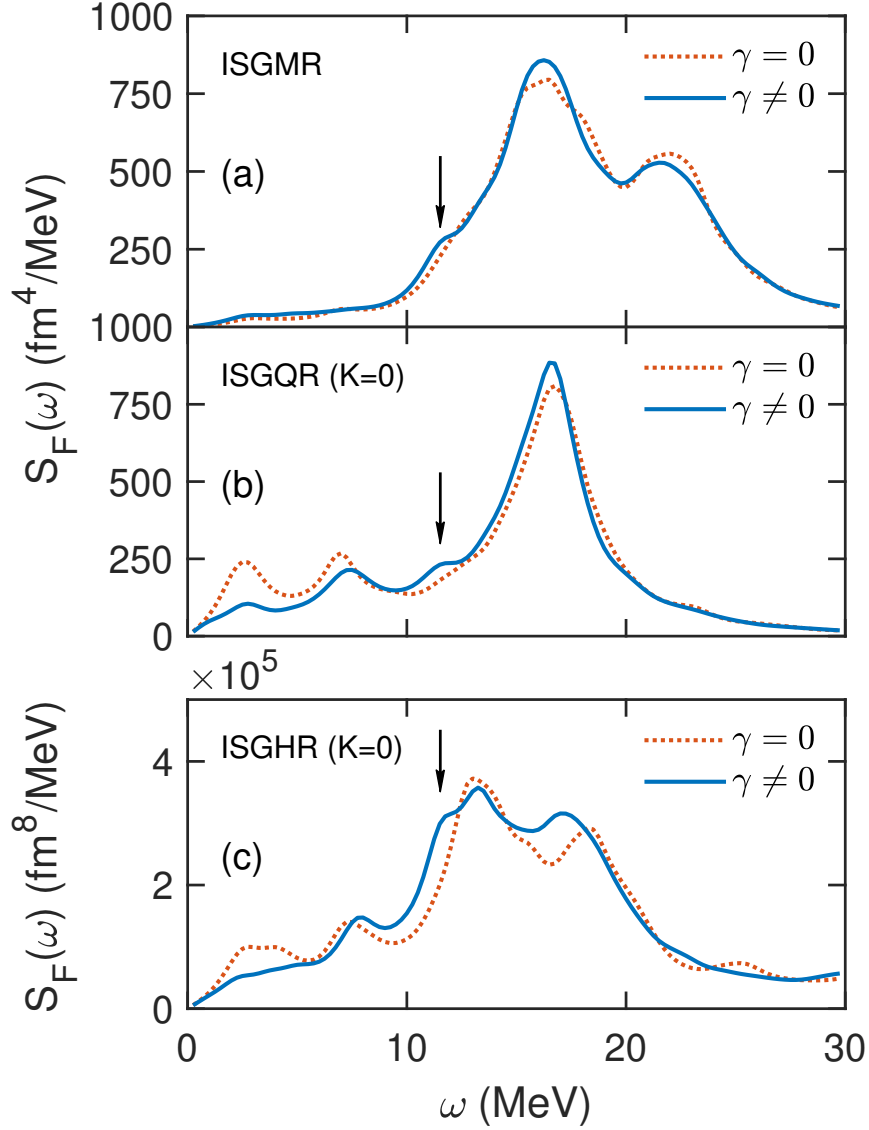}\\
	\caption{(a) ISGMR, (b) ISGQR with $K=0$, and (c) ISGHR with $K=0$ for $^{86}\textrm{Ge}$ with (solid line) and without (dotted line) the effect of triaxial deformation. The arrow in each panel denotes the peak due to the triaxial deformation.\label{Ge86_GMQHR}}
\end{figure}
In Fig. \ref{Ge86_GMQHR}, the strength functions of ISGMR, ISGQR with $K=0$, and isoscalar giant hexadecupole resonance (ISGHR) with $K=0$ for $^{86}\textrm{Ge}$ are presented. The strength functions are obtained by perturbing the nucleus with monopole, quadrupole, and hexadecupole vibrations, using the corresponding external field operators $\hat{Q}_{00}=r^2$, $\hat{Q}_{20} = r^2Y_{20}$, and $\hat{Q}_{40}=r^4Y_{40}$. A smearing width of 2 MeV is used to take into account the spreading effects \cite{Sun_PRC_2019_2}. The results for triaxially deformed case and for axially deformed case are drawn with solid lines and dotted lines, respectively. 

For the ISGMR, two branches of resonances can be identified when axial deformation is considered. 
The main peak of ISGMR locates higher and a pronounced peak appears at the lower excitation energy.
The monopole vibration has no directional projection, so that the ISGMR cannot split itself.
The reason for the splitting is  the monopole-quadrupole coupling between the ISGMR and the ISGQR with $K=0$  \cite{Niksic_PRC_2013,Kvasil_PRC_2016,Colo_PLB_2020}.
As can be identified in panel (a) and (b), the lower branch of ISGMR locates at the same position as the ISGQR with $K=0$. 
When the triaxial deformation is considered, the locations of the aforementioned peaks remain nearly unchanged, but the strength is slightly promoted for the lower one and reduced for the higher one. 
Because in triaxially deformed case, $^{86}\textrm{Ge}$ is more elongated in $z$-direction, thus the E0-E2 coupling is stronger and pumps more strength to the lower peak.

\begin{figure}	
	\centering
	%\includegraphics[width=0.45\textwidth]{./Figure_4.eps}\\	
	\includegraphics[width=0.7\textwidth]{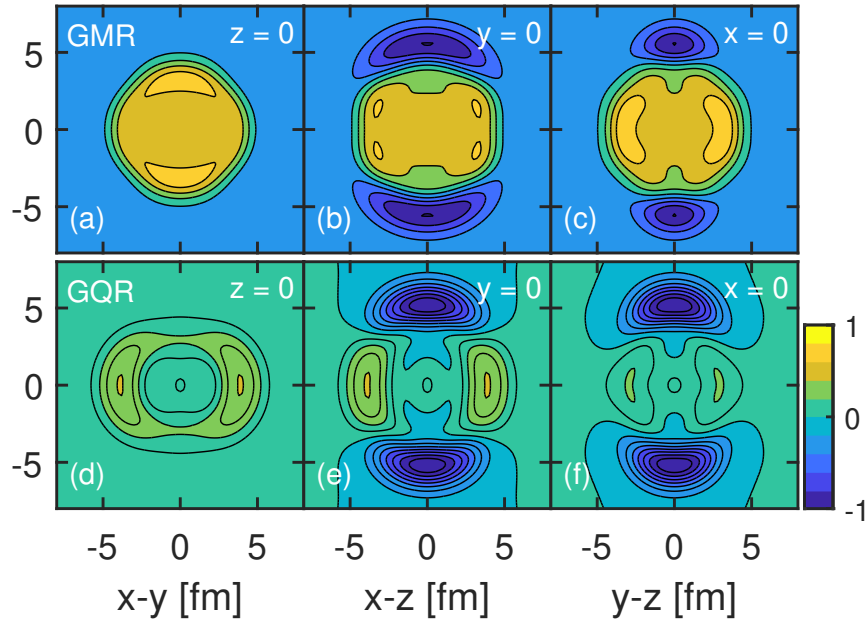}\\	
	\caption{Normalized transition densities at 16.5 MeV in $^{86}\textrm{Ge}$ for the resonance peaks of ISGMR [(a)-(c)], and the ISGQR with $K=0$ [(d)-(f)].\label{E0_E2}}
\end{figure}

Since the lower branch of the ISGMR is aroused by the E0-E2 coupling, it should contain considerable contributions from quadrupole vibrations.
To see this, in Fig. \ref{E0_E2}, the transition density, defined via
\begin{equation}
\delta\rho(\omega,x,y,z) = \sum_{i,j}\phi_i^{\dagger}(x,y,z)\delta\rho_{ij}(\omega)\phi_j(x,y,z),
\end{equation}
for the lower branch of ISGMR near 16.5 MeV is presented, 
with the distributions on the $z=0$, $y=0$, and $x=0$ planes in panel (a), (b), and (c).
The transition density for the ISGQR with $K=0$ are also presented in panel (d), (e), and (f) for comparison, from which it is clear to see that the vibration along the $z$-axis is out-of-phase to the vibration perpendicular to $z$-axis. The transition density for the ISGMR has a similar feature, but shows a strong mixture from the monopole vibration in the interior of the nucleus.
Therefore, triaxiality seems introducing no substantial difference to the main peaks of ISGMR, and the monopole-quadrupole coupling is still valid for the triaxially deformed nucleus $^{86}\textrm{Ge}$.

The most inspiring thing by scrutinizing the ISGMR in the triaxially deformed case is that, a small peak emerges near $\omega\approx11.5$ MeV in the strength function, as denoted by the arrow in Fig. \ref{Ge86_GMQHR}(a). The same resonance peak can also be found in the strength function of ISGQR presented in Fig. \ref{Ge86_GMQHR}(b). As for the ISGHR in  Fig. \ref{Ge86_GMQHR}(c), it appears as a shoulder at the low energy side of the main peak and enhances the strength function between 10 to 12 MeV. Indeed, triaxiality also impacts the low-energy strengths for ISGQR and ISGHR, e.g., quenching the strength functions around 2.5 MeV and 7 MeV.  However, as this work is devoted to investigate the effect peculiar to triaxiality, we would like to focus on the resonance peak at around 11.5 MeV which will vanish if the triaxiality is not considered. 
This triaxiality induced resonance peak is discernible for ISGMR, and is pronounced for ISGQR and ISGHR. 
The fraction of the energy weighted moment the triaxiality induced resonance peak exhausts, calculated by accumulating the energy weighted strength function from 10 to 12 MeV for the triaxially deformed case and then subtracting that for the axially deformed case, varies from 0.55\% for ISGMR, and 1.02\% for ISGQR, to 2.58\% for ISGHR. 

In a deformed nucleus, the monopole vibration may couple to the $K=0$ component of other even-$J$ vibrations. 
Since the resonance peak at 11.5 MeV responses to the monopole, the quadrupole, as well as the hexadecupole perturbations simultaneously, it should contains a mixture from the monopole, quadrupole, and hexadecupole vibrations. 
In close analogy to the ISGMR splitting that manifests the monopole-quadrupole (E0-E2) coupling, the triaxiality induced resonance peak might be an indicator to the coupling of the ISGMR and the ISGQR as well as the ISGHR with $K=0$, i.e., the monopole-quadrupole-hexadecupole (E0-E2-E4) coupling.

\begin{figure}	
	\centering
	%\includegraphics[width=0.45\textwidth]{./Figure_5.eps}\\	
	\includegraphics[width=0.7\textwidth]{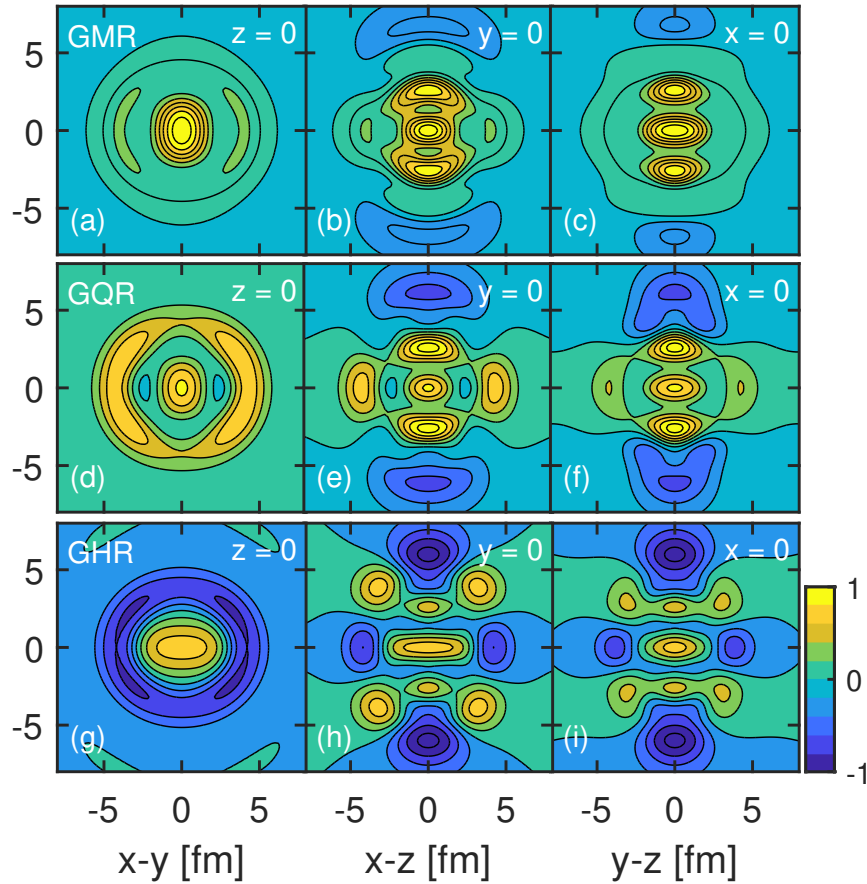}\\	
	\caption{Normalized transition densities at 11.5 MeV in $^{86}\textrm{Ge}$ for the resonance peaks of ISGMR [(a)-(c)], ISGQR with $K=0$ [(d)-(f)], and ISGHR with $K=0$ [(h)-(i)].\label{E0_E2_E4}}
\end{figure}
To verify the E0-E2-E4 coupling, and to check which vibration dominates, 
the microscopic structure of the triaxiality induced resonance peak at 11.5 MeV is analyzed in the following. 
The distributions of the transition density on the $z=0$, $y=0$, and $x=0$ planes are presented for ISGMR, ISGQR, and ISGHR in Fig. \ref{E0_E2_E4}. Since $^{86}\textrm{Ge}$ is of a triaxial shape with $\gamma=20.9^{\circ}$, the density distribution is stretched in $x$-direction and compressed in $y$-direction. 
Along the $z$-axis, the phase of the transition density in the interior region is different to those spread outside.
The transition densities show three nodes right next to each other in the interior region, which is a characteristic feature of hexadecupole vibrations. 
The three-nodes patter is prominent in the transition densities of ISGMR, ISGQR, and ISGHR for the triaxiality induced resonance peak, hence the contributions from hexadecupole vibrations are very strong and dominate. 
It should be noticed that, typically, for a quadrupole vibration, the phase of the transition density in the toroid on the $z=0$ plane is opposite to that in the top (bottom) lobe along $z$-axis; but for a hexadecupole vibration, they have the same phase.
Around the $z$-axis, according to different phases, both constructive and destructive interference of the monopole, quadrupole, and hexadecupole vibrations can be identified for the triaxiality induced peak in $^{86}\textrm{Ge}$.
For instance, the transition densities of ISGQR in panel (d), (e), and (f) show significant mixture between quadrupole and hexadecupole vibrations, 
with both in-phase and out-of-phase vibrations between the peripheral region and the interior region on the $z=0$ plane.
The transition density of ISGMR in panel (a), (b), and (c) is the result of the mixing of monopole, quadrupole, and hexadecupole vibrations.

In weakly bound neutron-rich nuclei, there may be low-lying resonance peaks which are ascribed to the excitation of excess neutrons \cite{Pei_PRC_2014,Timofeyuk_NPA_1993,Sun_PRC_2021,Sun_PRC_2022}. 
For $^{86}\textrm{Ge}$, affected by the pairing correlations, the 4 neutrons outside the $N=50$ magic core occupy the even-parity $sdg$ shell. 
Since ISGMR has an even parity and involves $2\hbar\omega$ particle-hole transitions \cite{Book_Harajeh}, in deformed case,
the lowest available state a neutron in $sdg$ shell can be excited to is $1i_{13/2}$, which is an intruder state to the $pfh$ shell thus has a desired even parity. 
In principle, $J=4$ transitions can be obtained through particle-hole configuration $2d_{5/2}\rightarrow 1i_{13/2}$, and $1g_{7/2} \rightarrow 1i_{13/2}$, etc. 
In axially deformed case, the single particle level with the angular momentum $\bm{j}$ splits into $(2j+1)/2$ two-fold degenerate orbits, those with larger absolute values of angular momentum projection $\Omega$ locate higher in energy when a nucleus is elongated. Though the particle-hole (ph) configurations $2d_{5/2} ~(\Omega_h = 5/2)\rightarrow 1i_{13/2} ~(\Omega_p = 1/2)$ and $1g_{7/2} ~(\Omega_h = 7/2)\rightarrow 1i_{13/2} ~(\Omega_p = 1/2)$ have energies close to 11.5 MeV
(which are 12.21 MeV and 8.99 MeV, respectively),
the transitions are forbidden by the selection rule $\Omega_p-\Omega_h=K=0$. While in the triaxially deformed case, the projection of the angular momentum $K$ of a nucleus is not a good quantum number. The aforementioned ph configurations are no longer hindered, which will contribute a $J=4$ component to the resonance peak near 11.5 MeV.

Indeed, the $J=4$ component of the vibration itself might be too weak to give noticeable effects alone. 
However, for the nucleus $^{86}\textrm{Ge}$, affected by the particle-hole configurations provided by the triaxiality, 
the hexadecupole vibrations interplay with monopole and quadrupole vibrations dramatically,
and eventually leads to a strong E0-E2-E4 coupling where hexadecupole vibration plays a dominate role.

\section{Conclusions}
The isoscalar giant resonances for the triaxially deformed nucleus $^{86}\textrm{Ge}$ are studied with the quasiparticle finite amplitude method based on the covariant density functional DD-ME2 and a separable pairing force. The ISGMR for $^{86}\textrm{Ge}$ splits into three components when the triaxial deformation is considered. The deformation induced double-peak structure for the ISGMR is clearly manifested when triaxiality is considered, so that the well-known monopole-quadrupole coupling is still valid for the triaxially deformed nucleus.  
In addition, a small resonance peak appears at $\omega\approx11.5$ MeV in the strength functions of the ISGMR when the triaxiality is taken into account. 
The triaxiality induced resonance peak also appears at the same energy in the strength functions of ISGQR, and ISGHR, implying the coupling among monopole, quadrupole, and hexadecupole vibrations.
Evidences for the monopole-quadrupole-hexadecupole coupling are found by analyzing the spatial distribution of the transition density. The contribution from hexadecupole vibration is enhanced by the triaxiality.
The emergence of the resonance peak at 11.5 MeV is a peculiar effect induced by the triaxiality. Therefore, it could serve as a fingerprint to identify the triaxiality of $^{86}\textrm{Ge}$ in future experiments.

\begin{acknowledgments}
This work is partly supported by the National Key Research and Development Program of China (Grants No. 2018YFA0404400 and No.2017YFE0116700), the National Natural Science Foundation of China (Grants No. 11621131001, No. 11875075, No. 11935003, and No. 11975031), the State Key Laboratory of Nuclear Physics and Technology, Peking University (Grant No. NPT2020ZZ01), and the China Postdoctoral Science Foundation under Grant No. 2020M680182. This work is supported by High performance Computing Platform of Peking University.
\end{acknowledgments}

\end{document}